\documentclass[conference]{cssconf}
\IEEEoverridecommandlockouts

\usepackage[utf8]{inputenc}
\usepackage[T1]{fontenc}
\usepackage[english]{babel}
\usepackage[dvipsnames]{xcolor}
\usepackage[noadjust]{cite}
\usepackage[linesnumbered,ruled]{algorithm2e}

\usepackage{amsmath,amssymb,amsthm,mathtools}
\usepackage{hyperref}
\usepackage{arydshln}
\let\labelindent\relax
\usepackage[inline]{enumitem}
\usepackage{cases}
\usepackage{tikz,pgfplots}
\usepackage{pgfgantt}
\usepackage{pdflscape}
\usetikzlibrary{calc,shapes,backgrounds,external}
\pgfplotsset{compat=newest}
\pgfplotsset{plot coordinates/math parser=false}
\tikzset{every picture/.style={line width=1pt}}
\pgfplotsset{every axis plot/.append style={line width=1pt}}
\tikzset{>=latex}

\SetKwInput{KwInput}{Input}                
\SetKwInput{KwOutput}{Output}              

\newtheorem{theorem}{Theorem}
\newtheorem{lemma}{Lemma}
\newtheorem{definition}{Definition}
\newtheorem{remark}{Remark}
\newtheorem{assumption}{Assumption}

\newlength\fheight
\newlength\fwidth

\def\arraystretchval{1.25}

\newcommand{\hop}{\mathsf{H} }

\title{\LARGE \bf Leveraging Non-Steady-State Frequency-Domain Data in~Willems'~Fundamental~Lemma}
\author{T.~J.~{Meijer}, M.~{Wind}, V.~S.~{Dolk}, and W.~P.~M.~H.~{Heemels}
\thanks{This research received funding from the European Research Council (ERC) under the Advanced ERC grant agreement PROACTHIS, no. 101055384.}%
\thanks{The authors are with the Control Systems Technology Section, Department of Mechanical Engineering, Eindhoven University of Technology, P.O. Box 513, 5600 MB Eindhoven, The Netherlands. {\tt\small t.j.meijer@tue.nl, m.wind@tue.nl, v.s.dolk@tue.nl, m.heemels@tue.nl}}}%

\begin{document}
\maketitle
\begin{abstract}
    Willems' fundamental lemma enables data-driven analysis and control by characterizing an unknown system's behavior directly in terms of measured data. In this work, we extend a recent frequency-domain variant of this result--previously limited to steady-state data--to incorporate non-steady-state data including transient phenomena. This approach eliminates the need to wait for transients to decay during data collection, significantly reducing the experiment duration. Unlike existing frequency-domain system identification methods, our approach integrates transient data without preprocessing, making it well-suited for direct data-driven analysis and control. We demonstrate its effectiveness by isolating transients in the collected data and performing FRF evaluation at arbitrary frequencies in a numerical case study including noise.
\end{abstract}

\section{Introduction}
Willems' fundamental lemma (WFL)~\cite{Willems2005} states that all finite-length input-output trajectories of an unknown linear time-invariant (LTI) system can be fully characterized using a single input-output trajectory, provided that the input sequence is persistently exciting (PE). This result has been instrumental in system identification through subspace methods~\cite{Verhaegen1992,McKelvey1996} and has more recently enabled significant advances in data-driven analysis and control. Notable applications of WFL include data-driven simulation~\cite{Markovsky2008}, state-feedback control~\cite{DePersis2020,Dorfler2023,Berberich2020b,vanWaarde2020}, and data-driven predictive control~\cite{Coulson2019,Berberich2021}. The success of WFL has also led to extensions for broader system classes, including linear parameter-varying systems~\cite{Verhoek2021}, descriptor systems~\cite{Schmitz2022}, stochastic systems~\cite{Faulwasser2023,Pan2022}, continuous-time systems~\cite{Rapisarda2023}, and various nonlinear systems~\cite{Berberich2020,Molodchyk2024}. For a more comprehensive overview, we refer the reader to~\cite[Table 1]{Faulwasser2023}. 

Recently, a frequency-domain version of Willems' fundamental lemma (FD-WFL) was introduced in~\cite{Meijer2024-nmpc,Ferizbegovic2021} and applied to data-driven analysis and control, including frequency-domain data-driven predictive control (FreePC)~\cite{Meijer2024-nmpc}. Unlike the original WFL and its associated data-driven methods, which rely on time-domain data, FD-WFL utilizes frequency-domain data, such as frequency-response-function (FRF) measurements, to characterize the behavior of an unknown system. This enables FD-WFL to exploit the abundance of frequency-domain data that is available after decades of working with classical frequency-domain (loop-shaping) control techniques, and to use the available expertise on collecting, interpreting and exploiting such data for control purposes. Importantly, FD-WFL achieves this \emph{without} the need to turn the frequency-domain data into a parametric (state-space) model. Another benefit of using frequency-domain data is that, compared to time-domain data, it is typically more dense in information, leading to smaller data sets, and we can work with easily-computable and interpretable uncertainty descriptions~\cite{Pintelon2012}. Opposed to time-domain WFL (TD-WFL), FD-WFL in its current form can deal with \emph{steady-state} data only. In practice, this limitation can be mitigated by pre-stabilizing the system if necessary and conducting sufficiently-long experiments in which transient effects have damped out and can effectively be ignored. While this approach--commonly used in practice for frequency-domain system identification, see, e.g.,~\cite{Pintelon2012}--is straightforward and often effective, it can be time-consuming and costly, especially for systems with slow time constants~\cite{Evers2020}. This challenge contrasts sharply with the increasing demands for higher productivity and throughput in many industrial applications, see, e.g.,~\cite{vanGerven2024}.

To overcome this limitation, we introduce an extension of FD-WFL that can effectively deal with transient phenomena. This extension, thereby, provides a rigorous mathematical framework for describing an unknown system's behavior using non-steady-state frequency-domain data, which can be used directly by many state-of-the-art data-driven analysis and control methods, such as, e.g., FreePC~\cite{Meijer2024-nmpc,Meijer2025-fd-wfl-arxiv}. Consequently, we are able to establish formal theoretical guarantees for these methods also when using non-steady-state data, which eliminates the need to wait for transients to decay and, thereby, significantly reduces experiment duration. To demonstrate their usefulness for data-driven analysis and control, we apply our results to data-driven FRF and transient evaluation, for which we also present a numerical case study. This application complements the FRF evaluation method based on time-domain data presented in~\cite{Markovsky2024}. 

In system identification literature, various methods have been proposed to handle transient frequency-domain data, including the local polynomial method (LPM)~\cite{Evers2020,Pintelon2010,Pintelon2012} and subspace identification methods~\cite{Cauberghe2006}. However, by embedding non-steady-state data directly within FD-WFL, our approach eliminates the need for additional processing steps, making it well-suited for direct data-driven analysis and control. 

The remainder of this paper is organized as follows. Section~\ref{sec:prelim} introduces some notation and relevant preliminaries. In Section~\ref{sec:problem}, we formalize the problem solved in the present paper. Next, Section~\ref{sec:transient-fd-wfl} presents an extension of the frequency-domain WFL to incorporate transient data, which constitutes our main contributions. Subsequently, Sections~\ref{sec:frf-eval} and~\ref{sec:case-study} present, respectively, an application of this result to data-driven frequency-response-function evaluation and a numerical case study thereof. Finally, we provide some conclusions in Section~\ref{sec:conclusions} and the proofs of our results in the Apppendix.

\section{Notation and preliminaries}\label{sec:prelim}
Let $\mathbb{R}$ denote the field of reals, $\mathbb{C}$  the complex plane, and $\mathbb{Z}$ the integers. We denote $\mathbb{Z}_{[m,n]}=\{m,m+1,\hdots,n\}$ and $\mathbb{Z}_{\geqslant m}=\{m,m+1,\hdots\}$, where $n,m\in\mathbb{Z}$ and $n\geqslant m$. The imaginary unit is denoted $j$, i.e., $j^2=-1$. For a complex-valued matrix $A\in\mathbb{C}^{n\times m}$, $A^\top$, $A^\hop$, and $A^*$ are its transpose, its complex-conjugate transpose, and its complex conjugate.

For any complex-valued function $v~\colon~\mathbb{Z}\rightarrow\mathbb{C}^{n_v}$, let \[v_{[r,s]}\coloneqq \begin{bmatrix} v_r^\hop & v_{r+1}^\hop & \hdots & v_{s}^\hop\end{bmatrix}^\hop\] denote the vectorized restriction of $v$ to the interval $\mathbb{Z}_{[r,s]}$. With some abuse of notation, we also use $v_{[r,s]}$ to refer to the sequence $\{v_k\}_{k\in\mathbb{Z}_{[r,s]}}$. Similarly, we use $0_{[r,s]}$ to refer to the length-$s-r+1$ sequence of null vectors of appropriate dimensions or their vectorized form \[0_{[r,s]}=\begin{bmatrix} 0 & 0 & \hdots & 0\end{bmatrix}^\top\] depending on the context. Finally, $0_{m\times n}$ denotes the $m$-by-$n$ zero matrix, $\otimes$ denotes the Kronecker product, and $\lceil\cdot\rceil$ the ceiling function.

\subsection{Persistence of excitation in frequency domain}
Let $\{\hat{\omega}_k^M\}_{k\in\mathbb{Z}_{[0,M-1]}}$ be the \emph{equidistant}\footnote{The preliminary results introduced in Section~\ref{sec:prelim} also apply to non-equidistant grids~\cite{Meijer2024-ecc-arxiv}.} frequency grid with 
\begin{equation}\label{eq:omega}
    \hat{\omega}_k^M = \frac{\pi k}{M}\text{ for all }k\in\mathbb{Z}_{[0,M-1]}.
\end{equation} 
Next, we recall the notion of persistence of excitation~\cite{Meijer2025-fd-wfl-arxiv,Meijer2024-nmpc} for the \emph{spectrum} $S_{[0,M-1]}$, with $S_k\in\mathbb{C}^{n_s}$ for all $k\in\mathbb{Z}_{[0,M-1]}$, obtained by sampling the discrete-time Fourier transform (DTFT) of a time-domain sequence $\{s_k\}_{k\in\mathbb{Z}}$ of $n_s$-dimensional vectors $s_k\in\mathbb{R}^{n_s}$, $k\in\mathbb{Z}$, at the frequencies $\hat{\omega}^M_{[0,M-1]}$, i.e.,
\begin{equation}\label{eq:DTFT}
    S_k = S(\hat{\omega}_k^M)\coloneqq \sum_{n\in\mathbb{Z}} s_ne^{-j\hat{\omega}_k^Mn}\text{ for all }k\in\mathbb{Z}_{[0,M-1]}.
\end{equation}
Since the sequence $\{s_k\}_{k\in\mathbb{Z}}$ is real-valued, $S(\omega)$ is symmetric, i.e., $S(-\omega)=S^*(\omega)$ for all $\omega\in\left[0,\pi\right)$. For $L\in\mathbb{Z}_{\geqslant 1}$, $m,n\in\mathbb{Z}_{\geqslant 0}$ with $n\geqslant m$, let $F_L~\colon~\mathbb{C}^{n_s(n-m+1)}\rightarrow\mathbb{C}^{n_sL\times (n-m+1)}$ be the function of $S_{[m,n]}$ given by
\begin{equation*}
    F_L(S_{[m,n]}) = \begin{bmatrix}
        W_L(e^{j\hat{\omega}^M_m})\otimes S_m & \hdots & W_L(e^{j\hat{\omega}^M_n})\otimes S_n
    \end{bmatrix},
\end{equation*}
where $W_L(z)\coloneqq \begin{bmatrix} 1 & z & \hdots & z^{L-1}\end{bmatrix}^\top$, $z\in\mathbb{C}$, and, similarly, let $\Psi_L~\colon~\mathbb{C}^{n_s(n-m+1)}\rightarrow\mathbb{C}^{n_sL\times (2(n-m)+1)}$ be the matrix-valued function of $S_{[m,n]}$ given by \[\Psi_L(S_{[m,n]}) = \begin{bmatrix} F_L(S_{[m,n]}) & F^*_L(S_{[m+1,n]})\end{bmatrix}.\] 

Using $F_L$, we recall the notion of PE for the complex-valued sequence $S_{[0,M-1]}$~\cite{Meijer2025-fd-wfl-arxiv,Meijer2024-nmpc}.
\begin{definition}\label{dfn:PE}
    The spectrum $S_{[0,M-1]}$ is said to be persistently exciting of order $L\in\mathbb{Z}_{[1,2M-1]}$, if the matrix $\Psi_L(S_{[0,M-1]})$ has full row rank.
\end{definition}
Observe that Definition~\ref{dfn:PE} exploits the symmetry of $S(\hat{\omega}_k^M)$ by including also $F_L^*(S_{[1,M-1]})$ in $\Psi_L(S_{[0,M-1]})$. Consequently, using $M$ frequencies, we can achieve up to $(2M-1)/n_s$ orders of PE, i.e., $L\leqslant (2M-1)/n_s$. 

\begin{remark}\label{rem:real}
    As detailed in~\cite{Meijer2025-fd-wfl-arxiv}, the conjugate symmetry in $\Psi_L(S_{[0,M-1]})$ allows it to be transformed to the real-valued matrix \[\begin{bmatrix}
        \mathrm{Re}(F_L(S_{[0,M-1]})) & \mathrm{Im}(F_L(S_{[1,M-1]}))
    \end{bmatrix},\]
    which can be beneficial for certain numerical solvers. Hence, Definition~\ref{dfn:PE} and other results in the sequel can be equivalently formulated in terms of real-valued matrices.
\end{remark}

\subsection{Frequency-domain Willems' fundamental lemma}
Next, we recall the frequency-domain version of WFL presented in~\cite{Meijer2025-fd-wfl-arxiv,Meijer2024-nmpc}. To this end, we consider discrete-time LTI systems, governed by
\begin{subnumcases}{\label{eq:system}\Sigma~\colon~}
    x_{k+1} &= $Ax_k + Bu_k,$\label{eq:system-state}\\
    y_k &= $Cx_k + Du_k,$\label{eq:system-output}
\end{subnumcases}
where $x_k\in\mathbb{R}^{n_x}$, $u_k\in\mathbb{R}^{n_u}$, and $y_k\in\mathbb{R}^{n_y}$ denote, respectively, the state, input, and output at time $k\in\mathbb{Z}$. The system~\eqref{eq:system}, i.e., the quadruplet of matrices $(A,B,C,D)$, is unknown. Let $\ell_{\Sigma}\in\mathbb{Z}_{[1,n_x]}$ denote the observability index of $\Sigma$, given by
\begin{equation*}
    \ell_{\Sigma} = \min\arg\max_{k\in\mathbb{Z}_{\geqslant 0}} \operatorname{rank} \mathcal{O}_k,
\end{equation*}
where $\mathcal{O}_k$ denotes the $k$-step observability matrix, i.e.,
\begin{equation}\label{eq:obs-mat}
\mathcal{O}_k = \begin{bmatrix} C^\top & (CA)^\top & \hdots & (CA^{k-1})^\top\end{bmatrix}^\top.
\end{equation}
\begin{assumption}\label{asm:ctrb}
    The pair $(A,B)$, is controllable.
\end{assumption}
Below, we formalize when two \emph{trajectories} $u_{[0,N-1]}$ and $y_{[0,N-1]}$ of length $N\in\mathbb{Z}_{\geqslant 1}$ are a solution to $\Sigma$.
\begin{definition}\label{dfn:input-output-trajectory}
    A pair of trajectories $(u_{[0,N-1]},y_{[0,N-1]})$ is called an input-output trajectory of $\Sigma$ in~\eqref{eq:system}, if there exists a state sequence $x_{[0,N-1]}$ satisfying~\eqref{eq:system-state} for $k\in\mathbb{Z}_{[0,N-2]}$ and~\eqref{eq:system-output} for $k\in\mathbb{Z}_{[0,N-1]}$. 
\end{definition}
Similarly, let us formalize \emph{steady-state} frequency-domain solutions to $\Sigma$, which are used as data in FD-WFL~\cite{Meijer2025-fd-wfl-arxiv}.
\begin{definition}\label{dfn:ss-input-output-spectrum}
    A pair of spectra $(U_{[0,M-1]},Y_{[0,M-1]})$ is called a steady-state input-output spectrum of $\Sigma$, if there exists a state spectrum $X_{[0,M-1]}$ satisfying
    \begin{equation}\label{eq:ss-spectra}
        \begin{aligned}
            e^{j\hat{\omega}_k^M}X_k &= AX_{k} + BU_k,\\
            Y_k &= CX_k + DU_k,
        \end{aligned}
    \end{equation}
    for all $k\in\mathbb{Z}_{[0,M-1]}$. In that case, the triplet $(U_{[0,M-1]},X_{[0,M-1]},Y_{[0,M-1]})$ is called a \emph{steady-state} input-state-output spectrum of $\Sigma$.
\end{definition}
Definition~\ref{dfn:ss-input-output-spectrum} does not account for any transient phenomena in~\eqref{eq:ss-spectra} and, therefore, these solutions are referred to as \emph{steady-state} input-(state-)output spectra. By eliminating $X_k$ from~\eqref{eq:ss-spectra}, it can be seen that Definition~\ref{dfn:ss-input-output-spectrum} includes the important special case where we are given data in the form of $M\in\mathbb{Z}_{\geqslant 1}$ FRF measurements $\{H(e^{j\hat{\omega}_k^M})\}_{k\in\mathbb{Z}_{[0,M-1]}}$ (see~\cite[Example 1]{Meijer2025-fd-wfl-arxiv}), where $H~\colon~\mathbb{C}\rightarrow\mathbb{C}^{n_y\times n_u}$ denotes the transfer function of $\Sigma$ given by \begin{equation}\label{eq:tf}
H(z)=C(zI-A)^{-1}B+D,\quad z\in\mathbb{C}.
\end{equation}

We are now ready to recall FD-WFL based on steady-state frequency-domain data~\cite{Meijer2025-fd-wfl-arxiv,Meijer2024-nmpc}, below.
\begin{lemma}\label{lem:ss-fd-wfl}
    Let $(\hat{U}_{[0,M-1]},\hat{X}_{[0,M-1]},\hat{Y}_{[0,M-1]})$ be a steady-state input-state-output spectrum of $\Sigma$ in~\eqref{eq:system} satisfying Assumption~\ref{asm:ctrb}. Suppose that $\hat{U}_{[0,M-1]}$ is PE of order $L+n_x$. Then, the following statements hold:
    \begin{enumerate}[label=(\roman*)]
    \item\label{item:ss-fd-wfl-1} The matrix \[\begin{bmatrix}
        \Psi_1(\hat{X}_{[0,M-1]})\\
        \Psi_L(\hat{U}_{[0,M-1]})
    \end{bmatrix}\] has full row rank;
    \item\label{item:ss-fd-wfl-2} The pair of trajectories $(u_{[0,L-1]},y_{[0,L-1]})$ is an input-output trajectory of $\Sigma$, if and only if there exist $G_0\in\mathbb{R}$ and $G_1\in\mathbb{C}^{M-1}$ such that
    \begin{equation*}\def\arraystretch{\arraystretchval}
\left[\begin{array}{@{}c@{}}
            u_{[0,L-1]}\\\hdashline[2pt/2pt]
            y_{[0,L-1]}
        \end{array}\right] = \left[\begin{array}{@{}c;{2pt/2pt}c@{}}
            F_L(\hat{U}_{[0,M-1]}) & F_L^*(\hat{U}_{[1,M-1]})\\\hdashline[2pt/2pt]
            F_L(\hat{Y}_{[0,M-1]}) & F_L^*(\hat{Y}_{[1,M-1]})
        \end{array}\right]
\left[\begin{array}{@{}c@{}}
            G_0\\
            G_1\\\hdashline[2pt/2pt]
            G_1^*
        \end{array}\right].
    \end{equation*}
    \end{enumerate}
\end{lemma}
An extension of Lemma~\ref{lem:ss-fd-wfl}, which is particularly useful when dealing with MIMO systems, that allows for the careful combination of multiple data sets, i.e., multiple input-output spectra, is also presented in~\cite{Meijer2025-fd-wfl-arxiv}.

\section{Problem statement}\label{sec:problem}
In general, Lemma~\ref{lem:ss-fd-wfl} requires infinitely-long measurements because the DTFT~\eqref{eq:DTFT} runs from $-\infty$ to $\infty$. This problem is also mentioned in~\cite{Meijer2025-fd-wfl-arxiv,Ferizbegovic2021,Meijer2024-nmpc}. In practice, this is often mitigated by waiting for transient phenomena to decay and, subsequently, collecting data that is (approximately) periodic. In this paper, we instead formulate an extension of FD-WFL that directly applies to non-steady-state data. This not only eliminates the need to wait for transients to decay during data collection, but it also provides a formal mathematical treatment of transient data.

To formally state this problem, we consider the discrete Fourier transform (DFT)
\begin{equation}\label{eq:DFT}
    S_k=S(\hat{\omega}_k^M)\coloneqq \sum_{n=0}^{2M-1} s_ne^{-j\hat{\omega}_k^Mn}\text{ for all }k\in\mathbb{Z}_{[0,M-1]},
\end{equation}
which, in contrast with~\eqref{eq:DTFT} on which Lemma~\ref{lem:ss-fd-wfl} is based, only requires a finite-length data sequence to compute. Here, we recall that $\{\hat{\omega}_k^M\}_{k\in\mathbb{Z}_{[0,M-1]}}$ are the equidistant frequencies in~\eqref{eq:omega}. Applying~\eqref{eq:DFT} to~\eqref{eq:system-state} yields
\begin{multline}\label{eq:development}
    AX_k + BU_k = \sum_{n=0}^{2M-1} x_{n+1}e^{-j\hat{\omega}_k^Mn}= \sum_{n=1}^{2M} x_ne^{-j\hat{\omega}_k^M(n-1)},\\
    = e^{j\hat{\omega}_k^M}X_k - x_0e^{j\hat{\omega}_k^M} + x_{2M}e^{-j\hat{\omega}_k^M(2M-1)},\\
    \stackrel{\eqref{eq:omega}}{=} e^{j\hat{\omega}_k^M}\left(X_k + x_{2M} - x_0\right),
\end{multline}
where $X_k=\sum_{n=0}^{2M-1} x_ne^{-j\hat{\omega}_k^Mn}$ and $U_k=\sum_{n=0}^{2M-1}u_ne^{-j\hat{\omega}_k^Mn}$, $k\in\mathbb{Z}_{[0,M-1]}$. This relation leads to the following notion of an input-output spectrum, in which the output equation is unchanged with respect to Definition~\ref{dfn:ss-input-output-spectrum}.
\begin{definition}\label{dfn:input-output-spectrum}
    A pair of spectra $(U_{[0,M-1]},Y_{[0,M-1]})$ is called an input-output spectrum of $\Sigma$, if there exists a state sequence $\{x_k\}_{k\in\mathbb{Z}_{[0,2M]}}$ such that the state spectrum $X_{[0,M-1]}$, with $X_k=\sum_{n=0}^{2M-1}x_ne^{-j\hat{\omega}_k^Mn}$ for $k\in\mathbb{Z}_{[0,M-1]}$, satisfies
    \begin{equation}\label{eq:input-output-spectrum}
        \begin{aligned}
            e^{j\hat{\omega}_k^M}X_k &= AX_k + BU_k + e^{j\hat{\omega}_k^M}(x_0-x_{2M}),\\
            Y_k &= CX_k + DU_k,
        \end{aligned}
    \end{equation}
    for all $k\in\mathbb{Z}_{[0,M-1]}$. In that case, the triplet $(U_{[0,M-1]},X_{[0,M-1]},Y_{[0,M-1]})$ is called an input-state-output spectrum of $\Sigma$.
\end{definition}      
Compared to the steady-state input-(state-)output spectra defined in Defintion~\ref{dfn:ss-input-output-spectrum},~\eqref{eq:input-output-spectrum} contains an additional term $e^{j\hat{\omega}_k^M}(x_0-x_{2M})$, which is not accounted for in Lemma~\ref{lem:ss-fd-wfl}. While $x_0$ and $x_{2M}$ appear in both~\eqref{eq:development} and Definition~\ref{dfn:input-output-spectrum}, they are assumed to be unknown in the sequel, i.e., our data consists only of the input-(state-)output spectra $(\hat{U}_k,\hat{X}_k,\hat{Y}_k)$ themselves. Observe that, for $2M$-periodic data, i.e., $x_0 = x_{2M}$, we recover Definition~\ref{dfn:ss-input-output-spectrum} as a special case. Moreover, by eliminating $X_k$ from~\eqref{eq:input-output-spectrum}, we find that the output data satisfies
\begin{equation*}
    Y_k = H(e^{j\hat{\omega}_k^M}) U_k + T(e^{j\hat{\omega}_k^M}),
\end{equation*}
where $H$ is the transfer function~\eqref{eq:tf} of $\Sigma$, and $T~\colon~\mathbb{C}\rightarrow\mathbb{C}^{n_y}$ is the transient given by
\begin{equation}\label{eq:transient}
    T(z) = C(zI-A)^{-1}z(x_0-x_{2M}).
\end{equation}

The objective of this paper is to characterize the behavior of the unknown system $\Sigma$ directly in terms of data, similar to TD-WFL and FD-WFL in Lemma~\ref{lem:ss-fd-wfl}, where the data consists of non-steady-state input-(state-)output spectra, as defined in Definition~\ref{dfn:input-output-spectrum}. We stress once more that $x_0$ and $x_{2M}$ are \emph{not} part of the data and, like the matrices $A$, $B$, $C$, and $D$, are unknown, making the approach in the sequel purely frequency-domain data driven. Additionally, we address the problem of data-driven transfer function and transient evaluation based on non-steady-state frequency-domain data. This enables us to separate the transfer function $H(z)$ in~\eqref{eq:tf} and transient $T(z)$ in~\eqref{eq:transient} contained in the collected data at any desired $z\in\mathbb{C}$.

\section{Willems' fundamental lemma using non-steady-state frequency-domain data}\label{sec:transient-fd-wfl}
In this section, we extend Lemma~\ref{lem:ss-fd-wfl} to incorporate non-steady-state data. To this end, we first introduce an augmented system in which we embed the transient and to which we, subsequently, apply Lemma~\ref{lem:ss-fd-wfl}.

\subsection{Augmented system}
To be able to deal with frequency-domain data consisting of general non-steady-state input-(state-)output spectra generated by $\Sigma$, we first introduce an augmented system $\tilde{\Sigma}$. This augmented system, which is inspired by~\eqref{eq:input-output-spectrum}, is governed by
\begin{subnumcases}{\label{eq:aug-system}\tilde{\Sigma}~\colon~}
    x_{k+1} &= $Ax_k + \tilde{B}v_k,$\label{eq:aug-system-state}\\
    y_k &= $Cx_k + \tilde{D}v_k,$\label{eq:aug-system-output}
\end{subnumcases}
where $v_k=(u_k,w_k)\in\mathbb{R}^{n_u+1}$ and
\begin{align*}
    \tilde{B}&\coloneqq \begin{bmatrix}
        B & x_0-x_{2M}
    \end{bmatrix},\\
    \tilde{D}&\coloneqq \begin{bmatrix}
        D & 0
    \end{bmatrix}.
\end{align*}
In defining $\tilde{\Sigma}$, we have embedded the transient phenomena into the $B$ matrix by considering $e^{j\hat{\omega}_k^M}$ as an additional input spectrum. Consequently, the augmented system $\tilde{\Sigma}$ has some useful properties, as stated below.
\begin{lemma}\label{lem:augmented-system}
    The solutions to $\tilde{\Sigma}$ in~\eqref{eq:aug-system} satisfy the following:
    \begin{enumerate}[label=(\roman*)]
        \item\label{item:aug-sys-1} The triplet of spectra $(U_{[0,M-1]},X_{[0,M-1]},Y_{[0,M-1]})$ is an input-state-output spectrum of $\Sigma$, if and only if $(V_{[0,M-1]},X_{[0,M-1]},Y_{[0,M-1]})$, with $V_k = (U_k,\Omega_k)$ and $\Omega_k=e^{j\hat{\omega}_k^M}$ for all $k\in\mathbb{Z}_{[0,M-1]}$, is a steady-state input-state-output spectrum of $\tilde{\Sigma}$;
        \item\label{item:aug-sys-2} The triplet of trajectories $(u_{[0,L-1]},x_{[0,L-1]},y_{[0,L-1]})$ is an input-state-output trajectory of $\Sigma$, if and only if $(v_{[0,L-1]},x_{[0,L-1]},y_{[0,L-1]})$, with $v_k = (u_k,0)$ for all $k\in\mathbb{Z}_{[0,L-1]}$, is an input-state-output trajectory of $\tilde{\Sigma}$.
    \end{enumerate}
\end{lemma}
The proof of Lemma~\ref{lem:augmented-system} can be found in the Appendix. In the next section, we will exploit these properties in order to develop an extension of Lemma~\ref{lem:ss-fd-wfl} that characterizes the behavior of the original system $\Sigma$ directly in terms of general (non-steady-state) input-(state-)output spectra.

\subsection{Willems' fundamental lemma}
Next, we will present our main result, which is a version of WFL that utilizes non-steady-state frequency-domain data. 
\begin{theorem}\label{thm:fd-wfl}
    Let $(\hat{U}_{[0,M-1]},\hat{X}_{[0,M-1]},\hat{Y}_{[0,M-1]})$ be an input-state-output spectrum of $\Sigma$ in~\eqref{eq:system} satisfying Assumption~\ref{asm:ctrb}. Suppose that $\hat{U}_{[0,M-1]}$ is such that $\hat{V}_{[0,M-1]}=\{(\hat{U}_k,\hat{\Omega}_k)\}_{k\in\mathbb{Z}_{[0,M-1]}}$, with $\hat{\Omega}_k = e^{j\hat{\omega}_k^M}$ for all $k\in\mathbb{Z}_{[0,M-1]}$, is PE of order $L+n_x$. Then, the following statements hold:
    \begin{enumerate}[label=(\roman*)]
        \item\label{item:fd-wfl-1} The matrix \[\begin{bmatrix}
            \Psi_1(\hat{X}_{[0,M-1]})\\
            \Psi_L(\hat{U}_{[0,M-1]})\\
            \Psi_L(\hat{\Omega}_{[0,M-1]})
        \end{bmatrix}\] has full row rank;
        \item\label{item:fd-wfl-2} The pair of trajectories $(u_{[0,L-1]},y_{[0,L-1]})$ is an input-output trajectory of $\Sigma$, if and only if there exist $G_0\in\mathbb{R}$ and $G_1\in\mathbb{C}^{M-1}$ such that
        \begin{equation*}\def\arraystretch{\arraystretchval}
    \left[\begin{array}{@{}c@{}}
                u_{[0,L-1]}\\
                0_{[0,L-1]}\\\hdashline[2pt/2pt]
                y_{[0,L-1]}
            \end{array}\right] = \left[\begin{array}{@{}c;{2pt/2pt}c@{}}
                F_L(\hat{U}_{[0,M-1]}) & F_L^*(\hat{U}_{[1,M-1]})\\
                F_L(\hat{\Omega}_{[0,M-1]}) & F_L^*(\hat{\Omega}_{[1,M-1]})\\\hdashline[2pt/2pt]
                F_L(\hat{Y}_{[0,M-1]}) & F_L^*(\hat{Y}_{[1,M-1]})
            \end{array}\right]
    \left[\begin{array}{@{}c@{}}
                G_0\\
                G_1\\\hdashline[2pt/2pt]
                G_1^*
            \end{array}\right].
        \end{equation*}
    \end{enumerate}
\end{theorem}
Theorem~\ref{thm:fd-wfl} provides a FD-WFL that can deal with general non-steady-state frequency-domain data of the form introduced in Definition~\ref{dfn:input-output-spectrum}. This is powerful because such data can be collected without having to wait for transients to decay. Theorem~\ref{thm:fd-wfl} can be directly used in many data-driven analysis and control methodologies, such as, e.g., FreePC~\cite{Meijer2024-nmpc}. In these methodologies, it enables us to provide theoretical guarantees also when using non-steady-state frequency-domain data, which is generally not possible using Lemma~\ref{lem:ss-fd-wfl} because any finite-length experiment will always contain some non-zero transient. In Section~\ref{sec:frf-eval}, we will demonstrate such an application to data-driven FRF and transient evaluation. Finally, we note that Theorem~\ref{thm:fd-wfl} can also be formulated in terms of real-valued matrices (see Remark~\ref{rem:real}).

Compared to the steady-state FD-WFL (Lemma~\ref{lem:ss-fd-wfl}), the FD-WFL in Theorem~\ref{thm:fd-wfl} requires additional data in the form of $\{\hat{\Omega}_k\}_{k\in\mathbb{Z}_{[0,M-1]}}$ (which we can compute directly from $\{\hat{\omega}^M_k\}_{k\in\mathbb{Z}_{[0,M-1]}}$) and a stronger PE condition on $\hat{V}_{[0,M-1]}=\{(\hat{U}_k,\hat{\Omega}_k)\}_{k\in\mathbb{Z}_{[0,M-1]}}$ instead of $\hat{U}_{[0,M-1]}$ alone. Note that $\hat{V}_{[0,M-1]}=\{(\hat{U}_k,\hat{\Omega}_k)\}_{k\in\mathbb{Z}_{[0,M-1]}}$ being PE of order $L$ requires that $2M-1 \geqslant L(n_u+1)$, while $\hat{U}_{[0,M-1]}$ itself being PE of order $L$ would only require that $2M-1\geqslant Ln_u$. In particular, we achieve this stronger PE condition by including $L^\prime\geqslant \lceil L/2\rceil$ additional frequencies in our data (i.e., increasing $M$ by $L^\prime$), and setting $\hat{U}_k=0$ at $L^\prime$ of those frequencies (while still measuring $\hat{Y}_k$) as also illustrated in Section~\ref{sec:case-study}. Interestingly, when dealing with non-steady-state time-domain data, TD-WFL does not require additional data nor a stronger PE condition. Compared to time-domain data, however, frequency-domain data is often denser in information and, as such, the use of non-steady-state frequency-domain data can still be more efficient than time-domain data.

\begin{remark}
Analogous to the extensions to multiple data sets presented in~\cite[Theorem 2]{vanWaarde2020} for time-domain data and~\cite[Theorem 2]{Meijer2025-fd-wfl-arxiv} for steady-state frequency-domain data, Theorem~\ref{thm:fd-wfl} can be extended to allow for multiple data sets. Doing so, enables us to utilize multiple input spectra $\hat{U}_k$ corresponding to the same frequency $\hat{\omega}_k^M$, $k\in\mathbb{Z}_{[0,M-1]}$, which is particularly useful when dealing with multi-input multi-output systems. In this case, it is generally not true that all data sets share common initial states $x_0$ and terminal states $x_{2M}$. However, this can be taken into account as follows. Let $E\in\mathbb{Z}_{\geqslant 1}$ denote the number of data sets, then the extension to multiple data sets can be obtained by applying~\cite[Theorem 2]{Meijer2025-fd-wfl-arxiv} to the augmented system $\tilde{\Sigma}$ with $v_k=(u_k,w_k)\in\mathbb{R}^{n_u+E}$ and \[\tilde{B} = \begin{bmatrix} B & x_0^1-x_{2M}^1 & \hdots & x_0^E-x_{2M}^E\end{bmatrix},\] where $x_0^e$ and $x_{2M}^e$ denote, respectively, the initial and terminal state corresponding to the $e$-th data set.

\end{remark}

\section{Frequency-response-function evaluation and~transient separation}\label{sec:frf-eval}
In this section, we use Theorem~\ref{thm:fd-wfl} to perform data-driven evaluation of the transfer function $H(z)$ and transient $T(z)$ at any $z\in\mathbb{C}$ that is not an eigenvalue of $A$. Additionally, we also isolate and evaluate the transient component $T(z)$ in~\eqref{eq:transient} in the collected data.
\begin{theorem}\label{thm:sep}
    Let $(\hat{U}_{[0,M-1]},\hat{Y}_{[0,M-1]})$ be an input-output spectrum of $\Sigma$ in~\eqref{eq:system} satisfying Assumption~\ref{asm:ctrb}. Let $L_0\in\mathbb{Z}_{\geqslant \ell_{\Sigma}}$ and suppose that $\hat{U}_{[0,M-1]}$ is such that $\hat{V}_{[0,M-1]}=\{(\hat{U}_k,\hat{\Omega}_k)\}_{k\in\mathbb{Z}_{[0,M-1]}}$, with $\hat{\Omega}_k=e^{j\hat{\omega}_k^M}$ for all $k\in\mathbb{Z}_{[0,M-1]}$, is PE of order $L_0+1+n_x$. Then, for any $z\in\mathbb{C}$ that is not an eigenvalue of $A$, the following statements hold:
    \begin{enumerate}[label=(\roman*)]
        \item\label{item:sep-1} For any sample $U_z\in\mathbb{C}^{n_u}$ of the input spectrum at $z$, the system of linear equations 
        \begin{multline}\label{eq:sep-1}\def\arraystretch{\arraystretchval}
            \left[\begin{array}{@{}c;{2pt/2pt}c@{}}
                0 & \Psi_{L_0+1}(\hat{U}_{[0,M-1]})\\
                0 & \Psi_{L_0+1}(\hat{\Omega}_{[0,M-1]})\\\hdashline[2pt/2pt]
                -W_{L_0+1}(z)\otimes I_{n_y} & \Psi_{L_0+1}(\hat{Y}_{[0,M-1]})
            \end{array}\right]\left[\begin{array}{@{}c@{}}
                Y_z\\\hdashline[2pt/2pt]
                G_Y
            \end{array}\right]\\ = \left[\begin{array}{@{}c@{}}
                W_{L_0+1}(z)\otimes U_z\\
                0\\\hdashline[2pt/2pt]
                0
            \end{array}\right]
        \end{multline}
        has a unique solution for $Y_z$, which is such that the pair $(U_z,Y_z)$ is a sample of a steady-state input-output spectrum of $\Sigma$ at $z$, i.e., $Y_z=H(z)U_z$;
        \item\label{item:sep-2} The system of linear equations 
        \begin{multline}\label{eq:sep-2}\def\arraystretch{\arraystretchval}
            \left[\begin{array}{@{}c;{2pt/2pt}c@{}}
                0 & \Psi_{L_0+1}(\hat{U}_{[0,M-1]})\\
                0 & \Psi_{L_0+1}(\hat{\Omega}_{[0,M-1]})\\\hdashline[2pt/2pt]
                -W_{L_0+1}(z)\otimes I_{n_y} & \Psi_{L_0+1}(\hat{Y}_{[0,M-1]})
            \end{array}\right]\left[\begin{array}{@{}c@{}}
                T_z\\\hdashline[2pt/2pt]
                G_T
            \end{array}\right]\\ = \left[\begin{array}{@{}c@{}}
                0\\
                W_{L_0+1}(z)\otimes z\\\hdashline[2pt/2pt]
                0
            \end{array}\right]
        \end{multline}
        has a unique solution for $T_z\in\mathbb{C}^{n_y}$, which corresponds to the transient~\eqref{eq:transient} present in the data, i.e., $T_z=T(z)$.
    \end{enumerate}
\end{theorem}
Theorem~\ref{thm:sep}.\ref{item:sep-1} allows us to evaluate the frequency-response-function of $\Sigma$ in the specific input ``direction'' $U_z$, which not only performs what is essentially an \emph{exact} interpolation of the data but also eliminates the transient $T(z)$. In fact, using Theorem~\ref{thm:sep}.\ref{item:sep-2}, we can also compute this transient separately. If we are interested in finding both $Y_z$ and $T_z$, we can also solve~\eqref{eq:sep-1} and~\eqref{eq:sep-2} simultaneously. Moreover, Theorem~\ref{thm:sep} extends~\cite[Proposition]{Meijer2025-fd-wfl-arxiv} to allow for non-steady-state frequency-domain data, and complements~\cite[Theorem 2]{Markovsky2024} for time-domain data. 

\section{Numerical case study}\label{sec:case-study}
In this section, we use Theorem~\ref{thm:sep} (and, thereby, Theorem~\ref{thm:fd-wfl}) to perform data-driven FRF and transient analysis based on non-steady-state frequency-domain data. To this end, we consider the benchmark of~\cite{McKelvey1996}, which is the fourth order single-input single-output system $\Sigma$ of the form~\eqref{eq:system} with transfer function 
\begin{multline*}
    H(z) = \\
    \frac{0.9626 z^4 + 0.4095 z^3 - 0.9718 z^2 + 0.26 z + 0.8618}{z^4 - 0.3306 z^3 - 0.5025 z^2 - 0.2347 z + 0.7925}.
\end{multline*}
Next, we consecutively consider noise-free data and data in which the output is corrupted by measurement noise.

\subsection{Noise-free data}
First, we consider the case with noise-free data, which we obtain by performing a multi-sine experiment with the $M=20$ frequencies in~\eqref{eq:omega}. We excite the $M/2=10$ odd frequencies, i.e., $\hat{U}_k=1$ for $k\in\mathbb{Z}_{[0,M-1]}^\mathrm{odd}\coloneqq\{1,3,\hdots,M-1\}$, and use the remaining even frequencies to do the transient estimation by taking $\hat{U}_k=0$ for $k\in\mathbb{Z}_{[0,M-1]}^{\mathrm{even}}\coloneqq \{0,2,\hdots,M-2\}$. It can be easily verified that the resulting spectrum  $\hat{V}_{[0,M-1]}=\{(\hat{U}_k,\hat{\Omega}_k)\}_{k\in\mathbb{Z}_{[0,M-1]}}$ of the augmented input is PE of order $L_0+1+n_x$ with $L_0=4\geqslant n_x=4$ so that we can apply Theorem~\ref{thm:sep}.

The resulting FRF and transient estimates obtained using Theorem~\ref{thm:sep}.\ref{item:sep-1} and Theorem~\ref{thm:sep}.\ref{item:sep-2}, respectively, are shown in Fig.~\ref{fig:noisefree-frf-eval}, along with the measured output spectrum $\hat{Y}_{[0,M-1]}$. Note that, for the even frequencies $k\in\mathbb{Z}_{[0,M-1]}^{\mathrm{even}}$, the measured output spectrum coincides precisely with the true transient, i.e., $\hat{Y}_k=T(e^{j\hat{\omega}^M_k})$ for $k\in\mathbb{Z}_{[0,M-1]}^{\mathrm{even}}$, because $\hat{U}_k=0$. For the odd frequencies $k\in\mathbb{Z}_{[0,M-1]}^{\mathrm{odd}}$, the measured output spectrum contains both the transfer function and the transient (and, thus, does not coincide with the true transfer function), i.e., $\hat{Y}_k=H(e^{j\hat{\omega}^M_k})+T(e^{j\hat{\omega}^M_k})$. It can be seen that the obtained estimates closely resemble the true FRF and the true transient. In fact, Fig.~\ref{fig:noisefree-frf-eval-error} shows the corresponding estimation errors, which are found to be close to machine precision. 

\begin{figure}[!tb]
    \setlength\fwidth{0.38\textwidth}
    \centering
    \input{images/noisefree-frf-eval.tex}
    \caption{Estimated (using Theorem~\ref{thm:sep}) FRF $Y_z$ (\kern0.5pt\protect\tikz[scale=0.6]{\protect\draw[color=blue,solid] (0,0.1)--(0.55,0.1);\protect\draw[color=white] (0,0)--(0.55,0)}) and transient $T_z$ (\kern0.5pt\protect\tikz[scale=0.6]{\protect\draw[color=red,solid] (0,0.1)--(0.55,0.1);\protect\draw[color=white] (0,0)--(0.55,0)}) of the system, for $z=e^{j\omega}$ with $\omega\in\left[0,\pi\right)$, based on the data $\hat{Y}_{[0,M-1]}$ split into odd (\kern0.5pt\protect\tikz[scale=0.3]{\protect\filldraw[color=blue,fill=white] (0,0) circle (0.25);}) and even (\kern0.5pt\protect\tikz[scale=0.3]{\protect\filldraw[color=red,fill=white] (0,0) circle (0.25);}) frequencies. The true transfer function $H(z)$ and transient $T(z)$ (\kern0.5pt\protect\tikz[scale=0.6]{\protect\draw[dashed] (0,0.1)--(0.55,0.1);\protect\draw[color=white] (0,0)--(0.55,0)}) are also depicted.}
    \label{fig:noisefree-frf-eval}
\end{figure}

\begin{figure}[!tb]
    \setlength\fwidth{0.38\textwidth}
    \input{images/noisefree-frf-eval-error.tex}
    \caption{Estimation errors $|H(z)-Y_z|$ (\kern0.5pt\protect\tikz[scale=0.6]{\protect\draw[color=blue,solid] (0,0.1)--(0.55,0.1);\protect\draw[color=white] (0,0)--(0.55,0)}) and $|T(z)-T_z|$ (\kern0.5pt\protect\tikz[scale=0.6]{\protect\draw[color=red,solid] (0,0.1)--(0.55,0.1);\protect\draw[color=white] (0,0)--(0.55,0)}) when using noise-free data.}
    \label{fig:noisefree-frf-eval-error}
\end{figure}

\subsection{Noisy data}
Next, we consider the case with noisy data, for which we modify the method based on Theorem~\ref{thm:sep} to include pre-processing of the data with the goal of approximating the noise-free data. To this end, we adopt the popular heuristic for observable systems, discussed in, e.g.,~\cite{Markovsky2024}, that noise-free data, for $L_0\geqslant \ell_{\Sigma}$ and under the appropriate PE conditions, satisfies
\begin{equation}\label{eq:heurestic}\def\arraystretch{\arraystretchval}
    \operatorname{rank}\left[\begin{array}{@{}c@{}}
        \Psi_{L_0+1}(\hat{U}_{[0,M-1]})\\
        \Psi_{L_0+1}(\hat{\Omega}_{[0,M-1]})\\\hdashline[2pt/2pt]
        \Psi_{L_0+1}(\hat{Y}_{[0,M-1]})
    \end{array}\right] = (n_u+1)(L_0+1) + n_x.
\end{equation}
This heuristic follows from Theorem~\ref{thm:fd-wfl}.\ref{item:fd-wfl-1} and the fact that 
\begin{multline*}\def\arraystretch{\arraystretchval}
    \left[\begin{array}{@{}c@{}}
        \Psi_{L_0+1}(\hat{U}_{[0,M-1]})\\
        \Psi_{L_0+1}(\hat{\Omega}_{[0,M-1]})\\\hdashline[2pt/2pt]
        \Psi_{L_0+1}(\hat{Y}_{[0,M-1]})
    \end{array}\right] = \\
    \def\arraystretch{\arraystretchval}\left[\begin{array}{@{}c;{2pt/2pt}c@{}}
        0 & I\\\hdashline[2pt/2pt]
        \mathcal{O}_{L_0+1} & \tilde{\mathcal{T}}_{L_0+1}
    \end{array}\right]\left[\begin{array}{@{}c@{}}
        \Psi_{1}(\hat{X}_{[0,M-1]})\\\hdashline[2pt/2pt]
        \Psi_{L_0+1}(\hat{U}_{[0,M-1]})\\
        \Psi_{L_0+1}(\hat{\Omega}_{[0,M-1]})
    \end{array}\right],
\end{multline*}
where \[\tilde{\mathcal{T}}_L\coloneqq\begin{bmatrix}
    \tilde{D} & 0 & \hdots & 0 & 0\\
    C\tilde{B} & \tilde{D} & \hdots & 0 & 0\\
    \vdots & \vdots & \vdots & \ddots & \vdots\\
    CA^{L-2}\tilde{B} & CA^{L-3}\tilde{B} & \hdots & C\tilde{B} & \tilde{D}
\end{bmatrix},\] and the observability matrix $\mathcal{O}_{L_0+1}$, given by~\eqref{eq:obs-mat}, is full column rank since $L_0\geqslant \ell_{\Sigma}$. We exploit~\eqref{eq:heurestic} in the following algorithm, which is inspired by~\cite[Algorithm 1]{Markovsky2024}.
\begin{algorithm}[!ht]
    \DontPrintSemicolon
    \label{alg:est}
    \caption{FRF and transient estimation.}
    
    \KwData{Input-output spectrum $(\hat{U}_{[0,M-1]},\hat{Y}_{[0,M-1]})$, frequencies $\hat{\omega}_{[0,M-1]}^{M}$ and model order $n_x$.}
    \KwInput{$U_z\in\mathbb{C}^{n_u}$ and $z\in\mathbb{C}$.}

    Let $L_0=n_x$.

    Compute singular value decomposition \[\left[\begin{array}{@{}c@{}}\def\arraystretch{\arraystretchval}
        \Psi_{L_0+1}(\hat{U}_{[0,M-1]})\\
        \Psi_{L_0+1}(\hat{\Omega}_{[0,M-1]})\\\hdashline[2pt/2pt]
        \Psi_{L_0+1}(\hat{Y}_{[0,M-1]})
    \end{array}\right] = \begin{bmatrix} \mathcal{U}_1 & \mathcal{U}_2\end{bmatrix}\mathcal{S}\mathcal{V}^\hop,\]
    with $\mathcal{U}_1\in\mathbb{R}^{(L_0+1)(n_u+1+n_y)\times (n_u+1)(L_0+1)+n_x}$.

    Solve the system \begin{multline*}\def\arraystretch{\arraystretchval}
    \left[\begin{array}{@{}c;{2pt/2pt}c@{}}
        \begin{bmatrix}
            0\\
            -W_{L_0+1}(z)\otimes I_{n_y}
        \end{bmatrix} & \mathcal{U}_1
    \end{array}\right]\left[\begin{array}{@{}c;{2pt/2pt}c@{}}
        Y_z & T_z\\\hdashline[2pt/2pt]
        G_Y & G_T
    \end{array}\right]\\ = \left[\begin{array}{@{}c;{2pt/2pt}c@{}}
        W_{L_0+1}(z)\otimes U_z & 0\\
        0 & W_{L_0+1}(z)\otimes z\\\hdashline[2pt/2pt]
        0 & 0
    \end{array}\right].\end{multline*}

    \KwOutput{$Y_z=H(z)U_z$ and $T_z=T(z)$.}    
\end{algorithm}

Algorithm~\ref{alg:est} uses the fact that there exists a singular value decomposition in which $\mathcal{U}_1$ and $\mathcal{U}_2$ are real, which is guaranteed by the following lemma.
\begin{lemma}\label{lem:structured-svd}
    Let $\mathcal{A} = \begin{bmatrix} \mathcal{A}_0 & \mathcal{A}_1 & \mathcal{A}_1^*\end{bmatrix}$ with $\mathcal{A}_0\in\mathbb{R}^{n\times m_0}$, $\mathcal{A}_1\in\mathbb{C}^{n\times m_1}$ and $m=m_0+2m_1$. Then, $\mathcal{A}$ admits a singular value decomposition $\mathcal{A}=\mathcal{U}\mathcal{S}\mathcal{V}^\hop$ with $\mathcal{U}\in\mathbb{R}^{n\times n}$ such that $\mathcal{U}^\top\mathcal{U}=I$, $\mathcal{V}=\begin{bmatrix} \mathcal{V}_0 & \mathcal{V}_1 & \mathcal{V}_1^*\end{bmatrix}^\hop$, $\mathcal{V}_0\in\mathbb{R}^{m\times m_0}$, $\mathcal{V}_1\in\mathbb{C}^{m\times m_1}$, such that $\mathcal{V}^\hop\mathcal{V} = I$, and \begin{equation}\label{eq:S}
    \mathcal{S}= \begin{bmatrix}\mathcal{S}_1 & 0_{r\times (m-r)}\\
    0_{(n-r)\times r} & 0_{(n-r)\times(m-r)}\end{bmatrix}\in\mathbb{R}^{(n-r)\times (m-r)},\end{equation}
    where $r=\operatorname{rank}\mathcal{A}$ and $\mathcal{S}_1$ is diagonal and positive definite.
\end{lemma}

As in the previous section, we perform a multi-sine experiment with the $M=20$ frequencies in~\eqref{eq:omega}, of which we excite the odd frequencies, i.e., $\hat{U}_k=1$ for $k\in\mathbb{Z}_{[0,M-1]}^{\mathrm{odd}}$, and use the remaining transient frequencies to do the transient estimation by taking $\hat{U}_k=0$ for $k\in\mathbb{Z}_{[0,M-1]}^\mathrm{even}$. The resulting spectrum $\hat{V}_{[0,M-1]}=\{(\hat{U}_k,\hat{\Omega}_k)\}_{k\in\mathbb{Z}_{[0,M-1]}}$ of the augmented input is PE of order $L_0+1+n_x$ with $L_0=4\geqslant n_x=4$. The measurements are obtained by measuring $100$ periods of the multi-sine, in which the output is corrupted by zero-mean Gaussian white noise with signal-to-noise-ratio of $20$ (i.e., $26.02$ dB), and computing the corresponding input-output spectrum. 

We use Algorithm~\ref{alg:est} to estimate the FRF and transient of the system, which yields the estimates shown in Fig.~\ref{fig:noisy-frf-eval} along with the measured output spectrum $\hat{Y}_{[0,M-1]}$. Despite the significant level of measurement noise, the obtained estimates remain close to the true transfer function and transient. In fact, the error remains below $-10$ dB for all frequencies, as seen in Fig.~\ref{fig:noisy-frf-eval-error} which shows the estimation errors. 

\begin{figure}[!tb]
    \setlength\fwidth{0.38\textwidth}
    \centering
    \input{images/noisy-frf-eval.tex}
    \caption{Estimated (using Theorem~\ref{thm:sep}) FRF $Y_z$ (\kern0.5pt\protect\tikz[scale=0.6]{\protect\draw[color=blue,solid] (0,0.1)--(0.55,0.1);\protect\draw[color=white] (0,0)--(0.55,0)}) and transient $T_z$ (\kern0.5pt\protect\tikz[scale=0.6]{\protect\draw[color=red,solid] (0,0.1)--(0.55,0.1);\protect\draw[color=white] (0,0)--(0.55,0)}) of the system, for $z=e^{j\omega}$ with $\omega\in\left[0,\pi\right)$, based on the noisy data $\hat{Y}_{[0,M-1]}$ split into odd (\kern0.5pt\protect\tikz[scale=0.3]{\protect\filldraw[color=blue,fill=white] (0,0) circle (0.25);}) and even (\kern0.5pt\protect\tikz[scale=0.3]{\protect\filldraw[color=red,fill=white] (0,0) circle (0.25);}) frequencies. The true transfer function $H(z)$ and transient $T(z)$ (\kern0.5pt\protect\tikz[scale=0.6]{\protect\draw[dashed] (0,0.1)--(0.55,0.1);\protect\draw[color=white] (0,0)--(0.55,0)}) are also depicted.}
    \label{fig:noisy-frf-eval}
\end{figure}

\begin{figure}[!tb]
    \setlength\fwidth{0.38\textwidth}
    \input{images/noisy-frf-eval-error.tex}
    \caption{Estimation errors $|H(z)-Y_z|$ (\kern0.5pt\protect\tikz[scale=0.6]{\protect\draw[color=blue,solid] (0,0.1)--(0.55,0.1);\protect\draw[color=white] (0,0)--(0.55,0)}) and $|T(z)-T_z|$ (\kern0.5pt\protect\tikz[scale=0.6]{\protect\draw[color=red,solid] (0,0.1)--(0.55,0.1);\protect\draw[color=white] (0,0)--(0.55,0)}) when using noisy data.}
    \label{fig:noisy-frf-eval-error}
\end{figure}

\section{Conclusions}\label{sec:conclusions}
In this paper, we extended the recently-introduced variant of Willems' fundamental lemma of~\cite{Meijer2024-nmpc,Meijer2025-fd-wfl-arxiv} to incorporate non-steady-state frequency-domain data. This advancement provides a formal mathematical framework for characterizing an unknown system's behavior using such data. Additionally, it significantly reduces measurement time by eliminating the need to wait for transients to decay. Importantly, Theorem~\ref{thm:fd-wfl} can be directly used in direct data-driven analysis and control methodologies, such as, e.g., FreePC~\cite{Meijer2024-nmpc}, to incorporate non-steady-state data. To illustrate this, we demonstrated its application in data-driven FRF and transient evaluation at arbitrary complex-valued frequencies, demonstrating the approach through a numerical case study. Future work will focus on incorporating noise in the analysis, and showing the effectiveness in the context of data-driven (predictive) control (FreePC).

\appendix
\section{Appendix}
\subsection{Proof of Lemma~\ref{lem:augmented-system}}
{\bf\ref{item:aug-sys-1}:} By Definition~\ref{dfn:ss-input-output-spectrum}, $(U_{[0,M-1]},X_{[0,M-1]},Y_{[0,M-1]})$ is an input-state-output spectrum of $\Sigma$, if and only if~\eqref{eq:input-output-spectrum} holds for all $k\in\mathbb{Z}_{[0,M-1]}$. This is equivalent to, for all $k\in\mathbb{Z}_{[0,M-1]}$,
\begin{align*}
    e^{j\hat{\omega}_k^M}X_k &= AX_k + \tilde{B}\begin{bmatrix}
        U_k\\
        e^{j\hat{\omega}_k^M}
    \end{bmatrix},\\
    Y_k &= CX_k + \tilde{D}\begin{bmatrix}
        U_k\\
        e^{j\hat{\omega}_k^M},
    \end{bmatrix},
\end{align*}
which is equivalent to $(V_{[0,M-1]},X_{[0,M-1]},Y_{[0,M-1]})$, with $V_k=(U_k,\Omega_k)$ and $\Omega_k=e^{j\hat{\omega}_k^M}$ for all $k\in\mathbb{Z}_{[0,M-1]}$, being a steady-state input-state-output spectrum of $\tilde{\Sigma}$. Hence, Lemma~\ref{lem:augmented-system}.\ref{item:aug-sys-1} holds.

{\bf\ref{item:aug-sys-2}:} By Definition~\ref{dfn:input-output-trajectory}, $(u_{[0,L-1]},x_{[0,L-1]},y_{[0,L-1]})$ is an input-state-output trajectory of $\Sigma$, if and only if, for all $k\in\mathbb{Z}_{[0,L-2]}$,
\begin{equation*}
    x_{k+1} = Ax_k + Bu_k = Ax_k + \tilde{B}\begin{bmatrix} u_k\\ 0\end{bmatrix},
\end{equation*}
and, for all $k\in\mathbb{Z}_{[0,L-1]}$, 
\begin{equation*}
    y_k = Cx_k + Du_k = Cx_k + \tilde{D}\begin{bmatrix} u_k\\ 0\end{bmatrix}.
\end{equation*}
This is equivalent to $(v_{[0,L-1]},x_{[0,L-1]},y_{[0,L-1]})$, with $v_k=(u_k,0)$ for all $k\in\mathbb{Z}_{[0,L-1]}$, being an input-state-output trajectory of $\tilde{\Sigma}$, whereby Lemma~\ref{lem:augmented-system}.\ref{item:aug-sys-2} holds. $\hfill\qed$

\subsection{Proof of Theorem~\ref{thm:fd-wfl}}
Let $(\hat{U}_{[0,M-1]},\hat{X}_{[0,M-1]},\hat{Y}_{[0,M-1]})$ be an input-state-output spectrum of $\Sigma$ satisfying Assumption~\ref{asm:ctrb}. Suppose that $\hat{U}_{[0,M-1]}$ is such that $\hat{V}_{[0,M-1]}=\{(\hat{U}_k,\hat{\Omega}_k)\}_{k\in\mathbb{Z}_{[0,M-1]}}$ with $\hat{\Omega}_{k}=e^{j\hat{\omega}_k^M}$ for all $k\in\mathbb{Z}_{[0,M-1]}$, is PE of order $L+n_x$. Note that, due to Assumption~\ref{asm:ctrb}, $\tilde{\Sigma}$ is also controllable. By Lemma~\ref{lem:augmented-system}.\ref{item:aug-sys-1}, $(\hat{V}_{[0,M-1]},\hat{X}_{[0,M-1]},\hat{Y}_{[0,M-1]})$ is a steady-state input-state-output spectrum of $\tilde{\Sigma}$. 

{\bf\ref{item:fd-wfl-1}:} Since $\hat{V}_{[0,M-1]}$ is PE of order $L+n_x$, it follows from Lemma~\ref{lem:ss-fd-wfl}.\ref{item:ss-fd-wfl-1} that \[\begin{bmatrix}
\Psi_1(\hat{X}_{[0,M-1]})\\
\Psi_L(\hat{V}_{[0,M-1]})
\end{bmatrix}\] has full row rank. Rearranging the rows yields Theorem~\ref{thm:fd-wfl}.\ref{item:fd-wfl-1}.

{\bf\ref{item:fd-wfl-2}:} It follows from Lemma~\ref{lem:ss-fd-wfl}.\ref{item:ss-fd-wfl-2} that the pair of trajectories $(v_{[0,L-1]},y_{[0,L-1]})$ is an input-output trajectory of $\tilde{\Sigma}$, if and only if there exist $G_0\in\mathbb{R}$ and $G_1\in\mathbb{C}^{M-1}$ such that \[\def\arraystretch{\arraystretchval}\left[\begin{array}{@{}c@{}}
    v_{[0,L-1]}\\\hdashline[2pt/2pt]
    y_{[0,L-1]}
\end{array}\right] = \left[\begin{array}{@{}c;{2pt/2pt}c@{}}
    F_L(\hat{V}_{[0,M-1]}) & F_L^*(\hat{V}_{[1,M-1]})\\\hdashline[2pt/2pt]
    F_L(\hat{Y}_{[0,M-1]}) & F_L^*(\hat{Y}_{[1,M-1]})
\end{array}\right]
\left[\begin{array}{@{}c@{}}
    G_0\\
    G_1\\\hdashline[2pt/2pt]
    G_1^*
\end{array}\right].\] Using Lemma~\ref{lem:augmented-system}.\ref{item:aug-sys-2} and by rearranging rows, we obtain Theorem~\ref{thm:fd-wfl}.\ref{item:fd-wfl-2}.$\hfill\qed$

\subsection{Proof of Theorem~\ref{thm:sep}}
Let $(\hat{U}_{[0,M-1]},\hat{Y}_{[0,M-1]})$ be an input-output spectrum of $\Sigma$ in~\eqref{eq:system} satisfying Assumption~\ref{asm:ctrb}. Let $L_0\in\mathbb{Z}_{\geqslant \ell_{\Sigma}}$ and suppose that $\hat{U}_{[0,M-1]}$ is such that $\hat{V}_{[0,M-1]}=\{(\hat{U}_k,\hat{\Omega}_k)\}_{k\in\mathbb{Z}_{[0,M-1]}}$, with $\hat{\Omega}_k=e^{j\hat{\omega}_k^M}$ for all $k\in\mathbb{Z}_{[0,M-1]}$, is PE of order $L_0+1+n_x$. 

By Lemma~\ref{lem:augmented-system}.\ref{item:aug-sys-1}, $(\hat{V}_{[0,M-1]},\hat{Y}_{[0,M-1]})$ is a steady-state input-output spectrum of $\tilde{\Sigma}$. The transfer function $\tilde{H}~\colon~\mathbb{C}\rightarrow\mathbb{C}^{n_y\times (n_u+1)}$ of $\tilde{\Sigma}$ is given by \begin{align*}
    &\tilde{H}(z) = C(zI-A)^{-1}\tilde{B}+\tilde{D},\\
     &\quad= \begin{bmatrix}
        C(zI-A)^{-1}B+D & C(zI-A)^{-1}(x_0-x_{2M})
        \end{bmatrix}.
\end{align*}
By~\cite[Proposition 2]{Meijer2025-fd-wfl-arxiv}, for any sample $V_z\in\mathbb{C}^{n_u+1}$ of the input spectrum at $z$, the system of linear equations 
\begin{multline}\label{eq:tf-eval-Sigma-tilde}\def\arraystretch{\arraystretchval}
    \left[\begin{array}{@{}c;{2pt/2pt}c@{}} 
        0 & \Psi_{L_0+1}(\hat{V}_{[0,M-1]})\\\hdashline[2pt/2pt]
        -W_{L_0+1}(z)\otimes I_{n_y} & \Psi_{L_0+1}(\hat{Y}_{[0,M-1]})
    \end{array}\right]\left[\begin{array}{@{}c@{}}
        \tilde{Y}_z\\\hdashline[2pt/2pt]
        G
    \end{array}\right]\\
    =\left[\begin{array}{@{}c@{}}
        W_{L_0+1}(z)\otimes V_z\\\hdashline[2pt/2pt]
        0
    \end{array}\right]
\end{multline}
has a unique solution for $\tilde{Y}_z$, which is such that the complex-valued pair $(V_z,\tilde{Y}_z)$ is a sample of the input-output spectrum of $\tilde{\Sigma}$ at $z$ (i.e., $\tilde{Y}_z=\tilde{H}(z)V_z$).

{\bf\ref{item:sep-1}:} By taking $V_z=(U_z,0)$ and rearranging rows, we find that $\tilde{Y}_z$ satisfies $\tilde{Y}_z=H(z)U_z$, and~\eqref{eq:tf-eval-Sigma-tilde} reduces to~\eqref{eq:sep-1}. Thereby, Theorem~\ref{thm:sep}.~\ref{item:sep-1} holds. 

{\bf\ref{item:sep-2}:} By taking $V_z=(0,z)$ and rearranging rows,  we find that $\tilde{Y}_z$ satisfies \[\tilde{Y}_z=C(zI-A)^{-1}z(x_0-x_{2M})\stackrel{\eqref{eq:transient}}{=}T(z).\]
Moreover,~\eqref{eq:tf-eval-Sigma-tilde} reduces to~\eqref{eq:sep-2}, whereby Theorem~\ref{thm:sep}.\ref{item:sep-2} holds.$\hfill\qed$

\subsection{Proof of Lemma~\ref{lem:structured-svd}}
Let $\mathcal{A} = \begin{bmatrix} \mathcal{A}_0 & \mathcal{A}_1 & \mathcal{A}_1^*\end{bmatrix}$ with $\mathcal{A}_0\in\mathbb{R}^{n\times m_0}$, $\mathcal{A}_1\in\mathbb{C}^{n\times m_1}$ and $m=m_0+2m_1$. Then, $\mathcal{A}\mathcal{A}^\hop$ is real and symmetric. Thus, there exists $\mathcal{U}\in\mathbb{R}^{n\times n}$ such that $\mathcal{U}^\top\mathcal{U}=I$ and $\mathcal{A}\mathcal{A}^\hop=\mathcal{U}\mathcal{S}^2\mathcal{U}^\top$ with $\mathcal{S}$ as in~\eqref{eq:S}, where $r=\operatorname{rank}\mathcal{A}$ and $\mathcal{S}_1$ is diagonal and positive definite. Let \[\mathcal{V} = \mathcal{A}^\hop \mathcal{U}\begin{bmatrix} \mathcal{S}_1^{-1} & 0\\ 0 & I\end{bmatrix}.\] Clearly, $\mathcal{U}\mathcal{S}\mathcal{V}^\hop = \mathcal{U}\begin{bmatrix}\mathcal{S}_1 & 0\\ 0 & I\end{bmatrix}\begin{bmatrix} \mathcal{S}_1^{-1} & 0\\ 0 & I\end{bmatrix}\mathcal{U}^\top\mathcal{A}=\mathcal{A}$ and \[\mathcal{V}^\hop\mathcal{V} = \begin{bmatrix} \mathcal{S}_1^{-1} & 0\\ 0 & I\end{bmatrix} \mathcal{U}^\top\mathcal{A}^\hop\mathcal{A}\mathcal{U}\begin{bmatrix} \mathcal{S}_1^{-1} & 0\\ 0 & I\end{bmatrix}=I.\] Finally, $\mathcal{V}$ has the desired structure $\mathcal{V}=\begin{bmatrix} \mathcal{V}_0 & \mathcal{V}_1 & \mathcal{V}_1^*\end{bmatrix}^\hop$ with
\begin{equation*}
    \mathcal{V}_0 = \begin{bmatrix}
        \mathcal{S}_1^{-1} & 0\\
        0 & I 
    \end{bmatrix}\mathcal{U}^\top A_0\in\mathbb{R}^{m\times m_0},
\end{equation*}
and 
\begin{equation*}
    \mathcal{V}_1 = \begin{bmatrix}
        \mathcal{S}_1^{-1} & 0\\
        0 & I 
    \end{bmatrix}\mathcal{U}^\top A_1\in\mathbb{C}^{m\times m_1}.
\end{equation*} 
$\hfill\qed$

\bibliographystyle{IEEEtran}
\bibliography{../../../bibliography/references}

\end{document}